# DeepTrust: A Deep Learning Approach for Measuring Social Media Users Trustworthiness

Majed Alrubaian[a], , Muhammad Al-Qurishi[a*], Sherif Omar[a] and Mohamed A. Mostafa[b]

[a] *Department of Software Engineering, College of Computer and Information Sciences, King Saud University, Riyadh, KSA*

[b] *Chair of Pervasive and Mobile Computing, Collage of Computer and Information Sciences, King Saud University, Riyadh, KSA*

Abstract – Veracity of data posted on the microblog platforms has in recent years been a subject of intensive study by professionals specializing in various fields of informatics as well as sociology, particularly in the light of increasing importance of online tools for news spreading. On Twitter and similar sites, it's possible to report on ongoing situations globally with minimal delay, while the cost of such reporting remains negligible. One of the most important features of this social network is that content delivery can be customized to allow users to focus only on news items covering subject matters they find interesting. With this in mind, it becomes necessary to create verification mechanisms that can ascertain whether the claims made on Twitter can be taken seriously and prevent false content from spreading too far. This study demonstrates an innovative System for verification of information that can fulfill the role described above. The System is comprised of four mutually connected modules: a legacy module, a trustworthiness classifier; a module managing user authority, and a ranking procedure. All of the modules function within an integrated framework and jointly contribute to an accurate classification of messages and authors. Effectiveness of the solution was evaluated empirically on a sample of Twitter users, with a strict 10-fold evaluation procedure applied for each module. The findings indicate that the solution successfully meets the primary objectives of the study and performs its function as expected.

## 1 Introduction

Online Social Networks (OSN) such as Twitter, Facebook, Instagram and YouTube have quickly been gaining popularity over the last decade, attracting new members at a tremendous rate. One of the major reasons for people to join such networks is easy access to real-time information in a convenient and cost-free fashion. Another key feature is the ability of members to express their views and opinions without any central authority to censor them [1-6]. However, one unintended consequence of free posting of content is the difficulty to differentiate between accurate and inaccurate factoids on social networks [2, 9]. Skilled users with malicious intent are often capable of disseminating false information, typically for personal gain or some other unsavory motive [3, 5, 9, 15]. As the majority of members become aware that the facts presented through the network may not be true, overall confidence in the platform is dramatically diminished. The answer to this problem is to study the trends and search for a solution that would be able to provide a quick evaluation of trustworthiness for each post. At present time, we are seeing a variety of strategies employed to this end [1, 2, 4, 7, 15,]. One group attempts to solve the issue by automatic classification that relies on artificial intelligence and recursive functions. An alternative approach is to combine human observations with machine-controlled processes, with the problem presented within the framework of cognition [3,6,8,10]. Other factors that were considered include analysis of the connection network and tracking high-authority users on social networks. Numerous software solutions were proposed for evaluating trustworthiness and displaying the results in a visual format, some of them in real time, through methods like graph comparison [16, 17]. Particular attention was paid to data credibility during dramatic occurrences like natural disasters or civil disturbances. The greatest obstacle to successful evaluation of data veracity on microblog platforms is the dynamic environment where new users are joining daily and large quantities of content are added every minute [2, 10, 15]. In general, the most significant problems in this field can be presented in the following way:

a) Size and dynamic nature of online media makes it difficult to pinpoint key factors that deserve to be studied in detail
b) Social networks are typically growing very fast, creating increasingly more connected matrixes of interactions where differentiating between actions of individuals becomes more challenging
c) Perceived trustworthiness of each network member is affected by a large number of inputs, including restructuring of his network of connections, actions of other members, displayed attitudes, etc.
d) Users with bad intentions can avoid detection by the existing anti-infiltration tools through several well-known loopholes. One way is for Twitter users to pay for acquiring contacts or to create large numbers of bot identities in order to disseminate the same message with slightly altered wording.
e) Assessing the effectiveness of any proposed solution is difficult to do with certainty due to the fact that research teams typically work with limited samples and lack the possibility to conduct large-scale testing.

Due to aforementioned factors, determining whether a certain network user (and his messages) can be trusted is an arduous and complicated process. With increased role of social networks in global news distribution to large populations, the importance of reliable methods for assessing trustworthiness of user-generated messages is rising. Our response involves a combined methodology optimized to recognize messages containing incorrect data and consequentially stop their further spreading. Major innovations proposed in this work and its basic structure can be expressed in the following manner:

a) We introduce DeepTrust a new approach for evaluating user trustworthiness based on recognition of a link between the author and the message, thus arriving to a more accurate estimation of data veracity. DeepTrust is composed of 3 mutually connected modules, each of which serves a specific function – tracking user reputation, a deep learning model for classification of trustworthiness scores and measuring user's authority about a topic. In combination, those modules are the backbone of the algorithm that can measure trustworthiness of individual social network messages.
b) We leverage user reputation to automatize the process of ranking authors on the merits of their knowledge and focus on a certain subject matter.
c) Each of the three modules was confirmed through a cross-validation procedure, using different data samples collected from the social network accounts. Verification outcomes imply that addition of user reputation as a determining factor has dramatically raised the accuracy of the user trustworthiness analysis in our solution.

Content of the research paper is organized in the following way: in Section 2 we provide an overview of existing solutions for determining the trustworthiness of user and content on microblog networks. In Section 3, specific challenges and the structure of our DeepTrust model are discussed. In Section 4, the solution for measuring trustworthiness of microblogs and their authors is laid out in detail. In Section 5, we summarize the outcomes of empirical testing of our solution. The paper ends with Section 6, which lists final observations and conclusions.

## 2  Related Work

In this section, current research on the topics of user trustworthiness and characteristics measurement are reviewed and presented in a brief form.

### 2.1  *Identifying misinformation on social platforms*

It is possible to classify all fake news discovery methods into two large groups based on the main source of the features [12]. That is why we can speak of news-based methods, and socially oriented methods [6]. The first group is characterized by the exploitation of verbal and visual cues as sources of relevant input. Different types of verbal content can be analyzed, from the writer's tone of voice and attention-seeking headlines, to morphology of the words and syntax of the sentences. On the other hand, visual cues are very valuable when the objective is to identify pictures that have been edited to distort the truth [13]. Socially oriented methods are less interested in the content of the message, and instead pay more attention to the profiles of users associated with dubious content, as well as other types of interactions, such as commenting on a post or expressing an emotion [13]. Informal networks of users can be analyzed as well, seeking for specific communities that assist in the propagation of inaccurate information through the social media platform or display common characteristics [21-23]. DeepTrust is proposed to include an examination of the user reputation and acquaintance, the content, and all social interactions based on it. A common trait of almost all feature-based methods is the narrow focus on leveraging user data for improving the performance of the classifier, rather than a thorough semantic analysis [22]. This study attempts to overcome this limitation and understand what each feature means and how it can be incorporated in a broader model aimed at recognition of misleading content.

*2.2    Evaluating the social media profiles of individual users*

Characteristics gleaned from user's social media activity can be conveniently divided into direct, which can be readily harvested from the user's profile in their original form, and indirect, which must be derived based on available data [15,23]. While direct features are far more commonly used, indirect ones can often be more helpful for understanding the user's patterns of activity. Some examples of indirect variables include gender, personality, or age group, so the methods for determining those characteristics are important for the overall improvement of the truthfulness of online news [24]. Authors in [26] suggested a promising method for guessing a user's gender based on textual content of his posts, while [27] tried incorporating usernames as features. They also tried to infer the personality type of the user based on linguistic factors. Similar approaches were devised for predicting the age group that the author of the posts likely belongs to [28,29], with a study by Schwartz combining age and gender prediction into a single model [30]. In this study, indirect characteristics are used alongside direct ones in order to maximize the discriminative ability of the model.

*2.3    Reputation-based Credibility assessment*

The question of information credibility on social networks has been extensively studied in the recent period, with various methods proposed for this purpose. Despite all the attention, at present time there are no decisively effective solutions that could provide adequate protection from misinformation and spam campaigns. Some of the most notable works include a study by Arubaian et al. [10] that was focused on Twitter and in many respects inspired this proposal. Gupta and Kumaraguru [32] elected to use the Support Vector Machine and Information Retrieval techniques to rank Twitter content based on its credibility. Meanwhile, Canini et al. [31] focused on establishing expertise for particular topics of conversation as a way of predicting credibility. All of the proposed methods can be described as either oriented towards machine learning algorithms (supervised or unsupervised) or relying on human evaluators to produce the rankings. Both groups have certain limitations – in the first case there is a need for massive processing power in order to obtain high accuracy, while in the latter there are practical challenges with distribution of questionnaires.

*2.4    Detection Arabic misinformation*

The topic of misinformation on the internet is heavily dependent on the chosen language, so works based in the English language can only partially guide research for Arabic content. One of such studies was conducted by Gupta et al. [32] and it proposes a semi supervised ranking model. However, there are specific studies that focus on the Arabic language specifically, with Sultan et al. [2] being the first work to directly deal with this problem. While this model works well for frequently discussed topics, it achieves poor results when used to rate news about ongoing events. Other classification models, like that developed by Al Mansour [33] have focused on extracting profile features (like presence of a profile picture) in order to determine reliability of the posted content. Some works were done in the area of sentiment extraction [2] providing sentiment lexicons for the Arabic language. More complex methods for determining sentiments of Arabic online users include work Al Sallab et al [35], but the authors of this study chose less demanding techniques. While this study was partially based on the aforementioned works, it addresses a gap related to Arabic-language Twitter content that wasn't adequately covered by earlier research.

## 3 DeepTrust Model Architecture

### 3.1 Design Overview

Structural design of the DeepTrust solution presented in this paper is given in Figure 1. There are 5 major building blocks that comprise it, that can be summarized in the following way: 1) Collection and archiving of messages; 2) Determining the trustworthiness score; 3) Determining the reputation score; 4) Evaluating the quality of user experience; and 5) Finding the overall trustworthiness measure, which can only be calculated with access to the results of the previous steps. Speaking in general, this five-step procedure can be considered an iterative mechanism that brings together the advantages of automated credibility testing and more nuanced trustworthiness measurements that rely on human judgement.

Data collection step was performed with the help of two separate Twitter API tools, one specialized for streaming content and another that identifies messages based on the topic. The first one was used to compile data samples for each particular event, while the other served to pick up the message histories of the involved users at the same time. Once the data was gathered on our servers, it was filtered and classified before it was presented for further use. The dataset obtained in this manner was separated into three distinct clusters – the messages about a certain topic, the users responsible for posting those messages, and the earlier message timelines of those users. Each cluster was utilized as input to a specific trust evaluation mechanism, providing indication whether the inspected element could be trusted or not.

The method for evaluating user reputation ignores the text of the message and instead focuses on analyzing the user's network of contacts as the foundation for its findings. Meanwhile, the credibility evaluation approach relies on access to checked facts, which can then be used for the training of a deep learning network, enabling the model to estimate the trustworthiness of new samples. User's knowledge about the topic is deduced from earlier messages, while the output of all three models is merged to come up with the final evaluation for each profile.

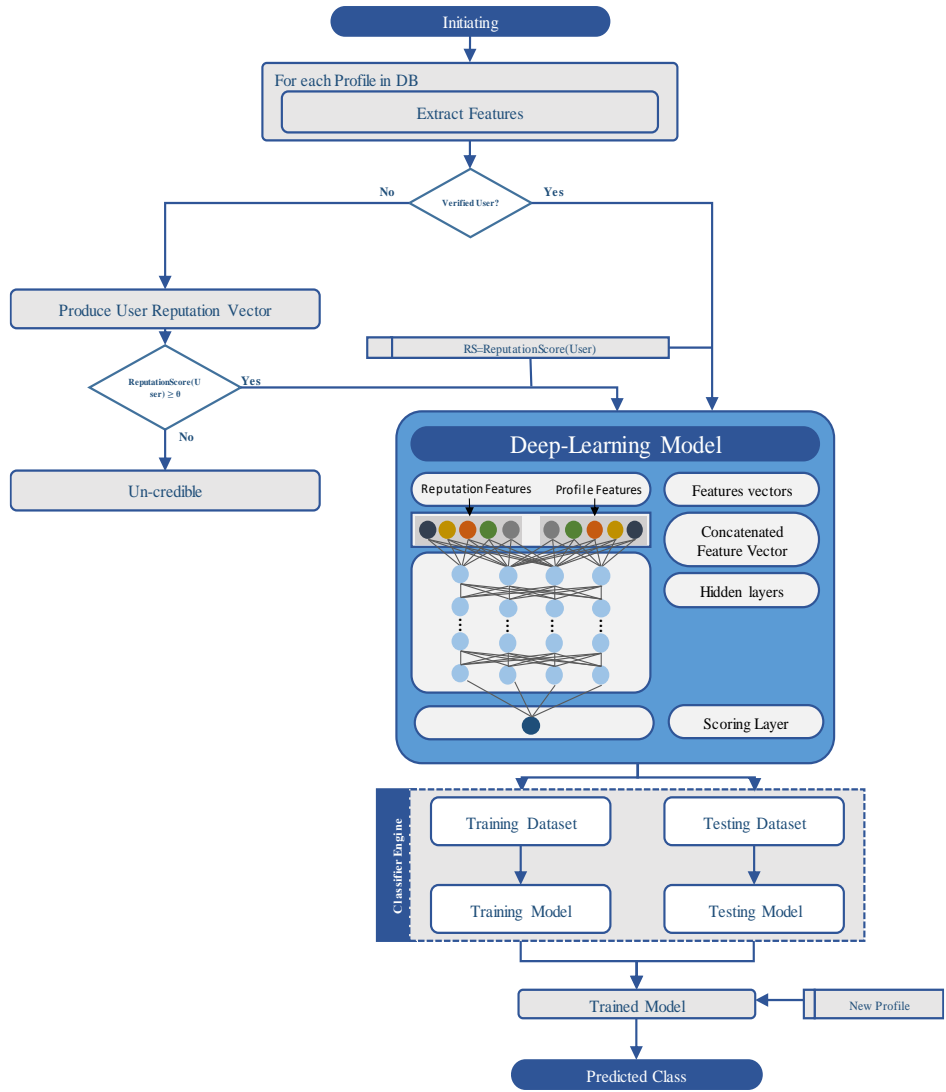

**Fig. 1.** DeepTrust Model Architecture

In this study, the main areas of interest are dataset preparation, as well as the statistical models for determining each aspect of trustworthiness, including reputation and user credibility.

### 3.2 Problem Statement

The purpose of the system is to determine how trustworthy a particular user on Twitter is, and whether the reputation of the person from whose account the tweet was sent is positive or negative. Given those objectives, it is very important that the key terms are used in a consistent and unambiguous way. In this work, the word "trustworthiness" pertains to the fact that the user is trusted and his profile contains truthful elements, and we use trustworthiness score as the relevant metric. On the other hand, the word 'reputation' expresses the typical reliability of the author of the message, while the associated score is the metric chosen for this purpose. However, it is important to differentiate between this feature and specific knowledge that the user might have about a given topic.

In a linguistic sense, trustworthiness is defined as a property related to being believed in, so it can be conceptualized as a combination of authority and persuasive ability. In particular, it pertains to the ability of an individual to cause others to believe his or her statements, which is a proactive property that can be exercised at will. If we are looking for synonymous words, the ones with the greatest semantic overlap would probably be 'believability', 'plausibility' and 'credibility, even if there are minor differences between them.

In the context of social networks, this word can mean several different things and be analyzed on multiple levels. Here is an overview of its usage in this study:

Use #1 – Message level trustworthiness – this quantity is mathematically expressed by $T(M)$ and it represents a quantitative summary of the assumed believability for each message. In turn, this indicates whether the message in question contains truthful information regarding a certain topic.

Use #2 – User level trustworthiness – this aspect of credibility is summarized as $T(U)$, and it quantifies how trustworthy the author of the tweet is within the social media platform. If this parameter is very low, that would indicate a high probability that the tweets coming from this author may contain inaccurate or misrepresented details.

Use #3 – User's reputation – this parameter is calculated exclusively on the basis of opinions of other users, although it relies on quantitative input. The value of user's reputation was obtained by analyzing the internal relationships, using variables such as follower ratio, replies, and retweets as proxies for measuring widespread respect in the community.

*3.3 Feature Engineering*

This section explains how the main features in our model were extracted from raw data, and divides them into three distinct groups based on their source as shown in Table 1:

*3.3.1 Message level*

Content features – those parameters are derived from the properties of the message content, for example the total number of characters in the message, total amount of responses, or the scope of re-publishing. In some cases the content may include special characters such as # signs or emojis, which can also have a significant impact on the perceived trustworthiness.

Emotional features – starting from a manually arranged list of affirmative and negative terms, we calculate the features that most successfully capture the sentimental tone of the conversation

*3.3.2 Account level*

A number of relevant data fields describing the user can be directly picked up from the profile page on the social network, such as his total number of Twitter contacts or the number of responses per post. On the other hand, personal information such as sex, educational background, political beliefs or personal preferences can only be deduced indirectly based on user's activity, and are considered 'hidden' features that require specific statistical procedures to obtain with certainty.

*3.3.3 Combined level*

In order to obtain most of the more expressive features, it is necessary to combine multiple message-level indicators and create new variables, for example the percentage of all messages containing a # sign or an external link, the mean sentiment value of messages, the number of repeated postings, etc. Features from this group were based on data collected on all three levels (content, author, event), with the combinations determined based on the assumed relevance for credibility evaluation. Topical relevance of the tweets was given serious consideration, while the accounts that had zero followers were eliminated from the calculation in order to preserve the integrity of the final values. There were two major reasons to include features obtained through hybrid means in the model. Some of the basic parameters could be easily manipulated, for example by artificially inflating the follower count through purchased contacts.

**4 Techniques for Capturing User credibility**

This section is dedicated to a detailed technical presentation of the approaches we selected to solve the central problem. That includes the methods for evaluating knowledge and popularity of every Twitter user, inspecting how trustworthy each message is, finding the most optimal features and ultimately determining whether any content units from the Twitter social network can be trusted or not. The overall structure of the model includes three major modules that are designed to work in unison, including the algorithm tasked with extracting the best features adopted from our previous work in [10], a module tasked with reputation tracking, a classification system for user credibility evaluation. In this chapter, each module is presented on a technical level and the main principles upon which it operates are explained.

## 4.1 Reputation Tracking

Understanding the concept of reputation on social networks and measuring it accurately are difficult problems that are further complicated by the presence of admiration for some users. This trend was experimentally confirmed by other authors, but the mechanisms of exerting social influence on Twitter are not completely known at this time. In response, we tried to design a solution that starts from the uniqueness of this platform and the type of messages found there. The biggest concern is to formulate the reputation metrics in such a way that their practical calculation is not overly resource consuming. While it may not be possible to find perfect metrics for Twitter, even a selection of partially useful indicators would be sufficient. Since reputation isn't explicitly tracked on Twitter, it was necessary to use other parameters such as sentimental reactions and popularity within the network to deduce this quantity based on them. The metric chosen to describe the reputation was mathematically represented as in equation 1 and equation 2 in section 4.3.2 and was computed for each network member.

### 4.1.1 Emotion Score

Emotional tone can significantly change how a certain person perceives the content of a message relating to some widely known issue. This is especially the case when said person already has a strong opinion and/or supports a specific position in the debate. When a message has implicit or explicit political connotation, it is possible that some participants in the conversation may have sinister motives to intentionally spread wrong or misinterpreted facts to advance their agendas. In this context, a sentiment encompasses several aspects of an interaction, including mental attitudes of all participants, their mutual relationships, as well as their pre-existing opinions. Another key component is deducing why certain network members are inclined to believe information coming from other members. A previous study that was focused on counting the affirmative and negative terms in a message determined a strong correlation between trustworthiness and positive sentiments, with very negative messages typically associated with the lowest level of credibility. We used Stanford tool for Arabic sentiment analysis[1] that was already demonstrated to be effective and deployed it to examine the sum of positive and negative messages in user's history and thus obtain the sentiment score for the whole profile.

### 4.1.2 Popularity Estimation

We also needed to define a simple way to estimate how popular a certain member was in the network and express this value in quantitative terms. The idea behind quantification was that such values could eventually serve for ranking purposes, allowing us to compare different Twitter users. All the indicators related to popularity were built into the algorithm, which was tasked with analyzing all messages coming from users who posted multiple tweets regarding a certain event. This allowed for the overall activity of a particular user in a defined time window to be concentrated and combined into a single quantity.

From practical perspective, it was discovered that republished messages (e.g., retweet), links to the users in other member's messages (e.g., reply and @mention), as well as following relationships are the steadiest contributing factors to this quantity (e.g., following and friendship). In essence, we posit that a user with huge of number of messages that were frequently re-broadcast by many users looks appealing to the network at large. Still, this doesn't negate the fact that the most important features are subjective in this case, so this is a serious obstacle to adequate evaluation of user's knowledge about relevant topics and his general level of popularity. This popularity matric known as the user acquittance which can be calculate by equation 1 as follows:

---

[1] https://nlp.stanford.edu/projects/arabic.shtml

$$ACQ_{scr} = \left(\frac{FWRcnt+RPLoth+MENoth+RTWcnt}{N}\right) \quad (1)$$

Where $FWRcnt$ is user follower count, $RPLoth$ is number of times he was replied by others, $MENoth$ is number of times he was mentioned by others, $RTWcnt$ is the number of time user tweets tweeted by others and $N$ is the number of users in a particular topic. The quantities defined above were then linearly combined to obtain the relative influence of a particular user on his network of contacts for a given topic, which can also be explained as his engagement which will be calculated by equation 2 as follows:

$$ACQ_{aff} = \begin{array}{l} \left(REP_{oth(N)} \times \sum_{i=1}^{N} \frac{ACQ_{scr(i)}}{REP_{oth(i)}}\right) \\ + \left(MEN_{oth(N)} \times \sum_{i=1}^{N} \frac{ACQ_{scr(i)}}{MEN_{oth(i)}}\right) \\ + \left(RTWcnt(N) \times \sum_{i=1}^{N} \frac{ACQscr(i)}{RTWcnt(i)}\right) \end{array} \quad (2)$$

Finally, all users were ranked according to their reputation, with all those whose score was lower than the threshold value set at $\theta \approx \{0, 0.1\}$ considered to be too unpopular to trust, while those ranked very highly were assumed to be sources that enjoy the greatest level of trust for the topic in question. Importantly, the ranking format was useful because it could be transferred to the next step and used for evaluation of general credibility. The techniques that start from user reputation have an advantage over other approaches and allow detection of high-impact users on the network with greater precision, regardless of the size and composition of the data sample. In addition, it is possible to move the threshold up or down and thus improve the efficiency of the process, minimizing the error rates during the prediction phase.

*4.2 User Trustworthiness Classification Model*

Considering the mass outreach of Twitter, the presence of inaccurate information can have a tremendous impact on the real world. It's a known phenomenon that certain people intentionally abuse this communication channel to distribute false information, often targeting certain companies or individuals perceived as competitors or threats. Typically, inaccurate content is designed to reflect the main objectives of the unscrupulous users and their primary tactics. Due to the difficult nature of ascertaining the facts in the online environment, some users may unwittingly help in the propagation of such content and voluntarily share it. Traditional media outlets are susceptible to this issue as well, and sometimes repeat unverified content from social networks with serious consequences. This is why a highly accurate system that could promptly indicate the level of trustworthiness for each Twitter user is sorely needed at this time but creating a tool that would meet both of those conditions at once is associated with many problems.

In our view, the discussion surrounding an issue can have a formative influence on the content of the online messages of the participants. There are multiple factors that can shape the discussion and impact the trustworthiness of the users and their posts, with political affiliation as one of the most obvious ones that are frequently encountered in online debates. American Elections represent one of the most divisive examples of this principle, where users were sharply divided into two camps and each side was propagating only the information that supported its position, usually by linking to sources sympathetic with their cause. Other issues tend to cause more uniform reactions, for example actions of Coronavirus disease (COVID-19) drew overwhelmingly negative sentiments among all Twitter users. Still, it's very common on social networks to doubt the veracity of certain messages and to publicly scrutinize the credentials of the users.

The central idea behind this element was to instantly ascertain whether a certain social media user is trustworthy or not. To this end, we propose a deep neural network with direct human supervision, aiming to maximize recall of the system and keep false positives and false negatives as low as possible. The deep learning model consists of seven layers as shown in figure 2.

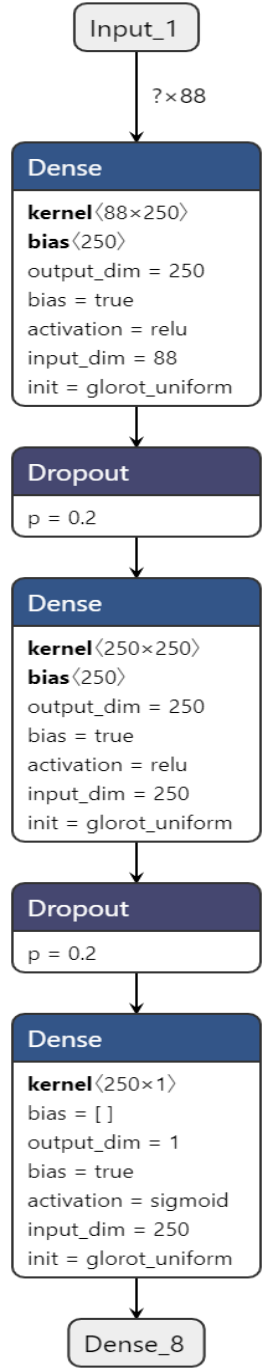

**Fig. 2.** Seven Layers of the Deep Learning Model

    The first layer in the model called the input layer which contains the input features. It has a dimension of $(k,1)$ where $k$ represents the length of the feature vector $F = (f_1, f_2, f_3, \ldots, f_k)$. Each feature will be assigned to an activation function and passed it on to the next layer in the network. A simple activation function can be shown in equation 3 as follows:

$$g = ReLU(A^T \times F + b) \qquad (3)$$

Where, $A^T, b$ are the weight and bias parameters associated with the input features. Each node in each layer has two steps of calculations. The first step is calculating the value of $(A^T \times F + b)$ then calculating the Rectified Linear Unit *ReLU* of the output of the first step.

The second layer in the model is a dense layer with two diminution (*k*, *h*) where *h* represents the number of hidden units which is here 250 nods. Similarly, each node will generate an activation value that will be passed on to the subsequent layer and so on. We have added a dropout layers for optimizations purpose [37]. By applying samples from a Bernoulli distribution, we have randomly dropped some of the hidden units by zeroing their activation with probability *p* during the training phase. The outputs of this layer are bounded by $\frac{1}{1-p}$. The fourth layer is a dense layer that has two dimensions (*h*, *h*) followed by another dropout layer. The last dense layer will have a dimension of (*h*, 1) where *h* is the input dimension and 1 is the output dimension. The used activation function is sigmoid as shown in equation 4.

$$\sigma(x) = \frac{1}{1+e^{-x}} \qquad (4)$$

Finally, the output layer is a single node layer since we are using a binary classification where the label $y \in \{trusted, not\text{-}trusted\}$.

There were two starting assumptions that can be formalized in this way:
- Credibility value of each user found on Twitter can be quickly and effectively assessed with our method
- The selected set of features bears crucial significance for the accurate classification of content

The full list of features that were chosen to evaluate trustworthiness of Twitter users was presented in an earlier in section 3.5 along with explanations about their collection methods. A majority of those parameters are specific to the social media platform and would not be available or suitable for use in a different context.

It was necessary to use a human expert to create the ground truth input, which could then be used to train the classification engine and improve its effectiveness on the credibility evaluation task. Training was conducted in a supervised fashion, with the number of classes to be differentiated between set at two. An assumption is made that the impact of each feature is not dependent on any other features, allowing for the calculation of combined probabilities for the entire set of features.

## 5 Experimental Results

Effectiveness of the proposed model is evaluated through several steps in which a collection of real users' profiles data was used. The following sections will explain that in details.

### 5.1 Data Sample

The method suggested in the earlier sections was used with realistic Twitter data pertaining to the American elections. The entire sample consisted of 6541 unique profile with nearly 14.3 million messages, and it was divided in into three datasets: 80% as training set, 10% as validation and 10% as testing dataset. Each of those samples includes several million tweets. Due to the rules of the platform, a maximum of just over 3,200 messages could be collected from a single author.

### 5.2 Data processing

All users in the sample were annotated as either trust or not trust, before proceeding to the next step in which various features were derived from the data. Since Twitter allows only access to data about users that are sent to the entire network and not about links between users, it was impossible to create a graph illustrating social connections. Because of this, some of the features reliant on spatial representation of connections between users had to be omitted, for example centrality and distance based metrics. Despite the considerable potential of such variables, difficulty of access was the main reason why they were not selected for classifier training. In general, features that could be obtained without excessive effort were prioritized, as the idea was to build a practical and easily executable model. All introductory messages, Tweets, or users that repeated existing content were removed from the sample, aiming to secure that every user is topically focused, which was the final step in a complex and nuanced procedure of data augmentation. The final list of selected features is provided in a previous section, and they can be classified in three major groups – message-oriented, author-oriented, and mixed. All the data was examined based on two main criteria, namely the semantic theme of the message, and earlier activity of the message's author.

*5.3   Profile Activity Analysis*

Profile data of some users was blindly chosen from the sample in order to inspect their earlier activities on the network. For this purpose, cumulative distributive function was calculated, with one example from each group. The results of this procedure clearly show that users with low trustiness are typically composed of fewer total characters in comparison with trust users. Two groups also differ regarding the dominant sentiment, as messages with low credibility are more frequently written in a negative tone, in contrast to highly trustworthy. A similar trend is observed with the number of links with other users, with senders of dubious content followed by significantly fewer people than those who more frequently share credible content. While this study confirmed the ability of the tested features to aid in the process of recognition of false information, they are only a small subset of all relevant features and the list could potentially be complemented with many other parameters.

*5.4   Relationships between Variables*

Each of the selected features must be understood in combination with other relevant factors that might be modifying its influence. To this end, we used a visual method and examined the connections between some of the variables included in our mixed model. Legitimate and false users were represented as color-coded points, and scattering plots were created for each variable where groupings of points of certain color indicated whether the feature is relevant and to which extent. To make an illustration, it was immediately obvious from the plots that users with listed count near to zero were far more likely to be evaluated poorly, and the same was apparent for users that do not mention any other user profiles.

On the other hand, messages sent by users who have wide networks of followers and those that rarely use negative expressions were overwhelmingly seen as completely legitimate. Interestingly, the shape of the plot could also readily signal that the inspected variable is not correlated with trustworthiness.

*5.5   Training the Neural Classifier*

The algorithm for classification of tweets according to their credibility requires a training stage before it can be effectively deployed. For best results, the model should be trained on a dataset that contains numerous examples of both types of users. The collection of users with low credibility was compiled by establishing external references for truthfulness and identifying the authors/messages who do not abide by the truth. Because of this, all users regardless of their conents must be examined with the neural classifier. The profiles with no social connections were excluded from the sample, while a special role was reserved for individuals with relevant credentials for certain topics, as their expertise was used as a source of information for data labeling. The people tasked with labeling had previous experience with Twitter, and were instructed to search all publically available sources before making the final decision regarding truthfulness of a user. By gaining exposure to examples of both content types and to the constellations of various features, the neural model gradually becomes more capable of predicting the truthfulness level of unseen samples.

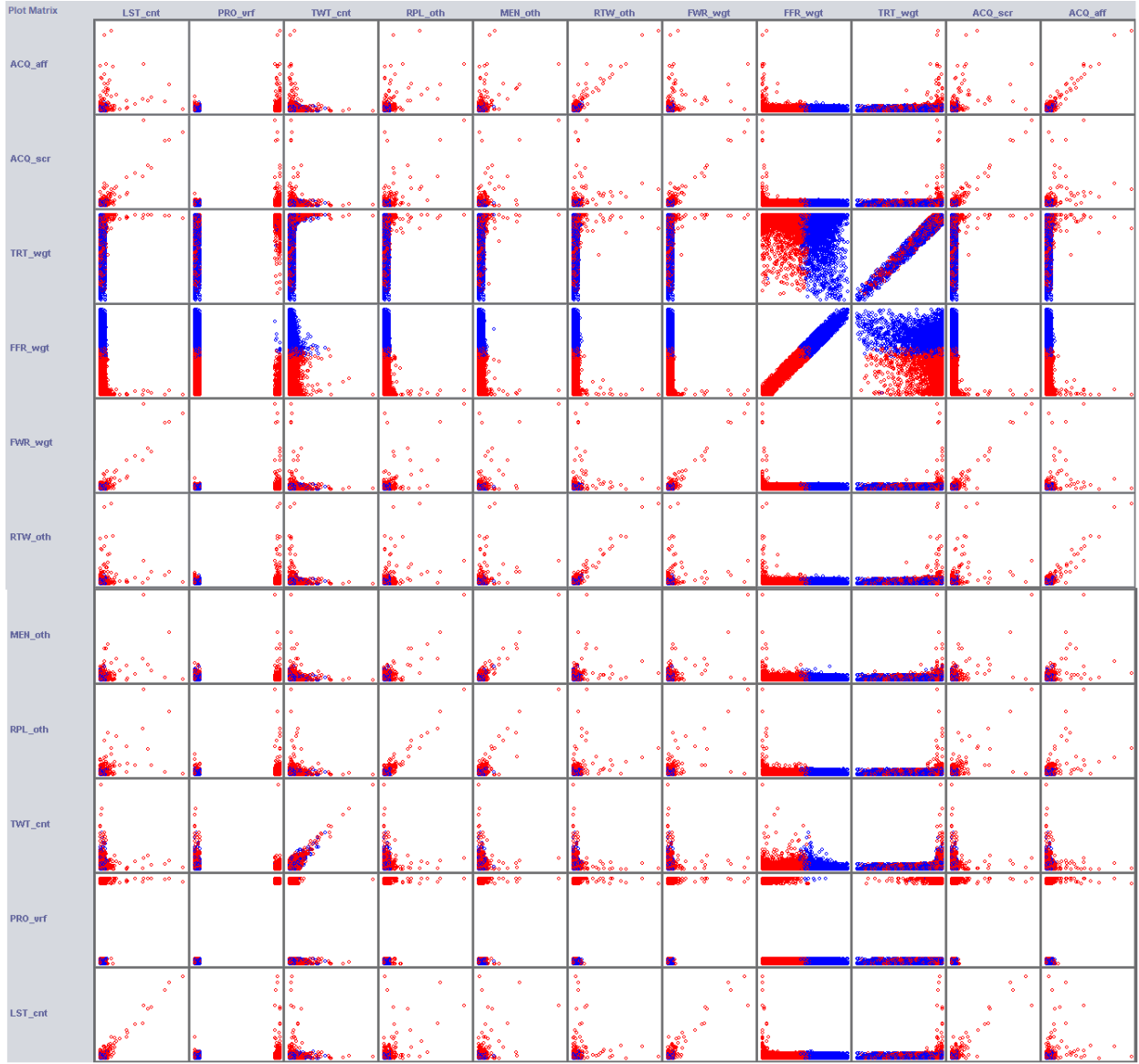

**Fig. 3.** Matrix Scatter Plots for Selected Variables

### 5.6 Evaluation of the model's effectiveness

After completing our fundamental algorithm based on user's reputation, we tested its effectiveness against several different neural classifiers. Those classifiers were trained using the annotated data sample, and were both trained and evaluated with a 10-fold statistical procedure. On the other hand, the reputation-based method identified very small portions (less than 0.01) of the dataset as completely credible, and those users were treated as reliable sources of information. Optimization was performed with the idea to improve accurate recognition for both classes, which necessitated an operation intended to minimize the loss function:

$$\varepsilon_m = \min EL(d, f_m(d))$$

In this formula, d denotes any network item (profile or message) that needs to be classified, while $f_m(d)$ represents the most optimal classification function. The validation procedure consists of several stages, the first of which includes randomization of the sample. Immediately following this step, the sample is divided into ten equally sized units, which are used as independent datasets. The final step includes running the same number of cycles, with each cycle using one of those sub-samples as input for testing, while the remaining samples are used for model training. Once all ten cycles are executed, average values are calculated to come up with the final result. Error is an important factor in this model, and can be calculated by the following formula:

$$\varepsilon_m = \frac{1}{K} \sum_{i=t}^{k} \varepsilon_m(i)$$

The value of m can be selected in such a way to keep the error at the minimum. Once the best value of m is found, it can be used to once again train the classifier with all of the sample subsets. The chosen validation sequence is superior to all alternatives (i.e. blind sample division) for multiple reasons, including the fact that it can ensure that no part of the sample is repeated during training. On the other hand, no part of the sample is left unused either, so the cumulative effect is that error levels can be evaluated much more precisely. Each of the cycles is completely independent from all the others, which further improves reliability of the results.

There were four different classification algorithms used for evaluation – decision tree algorithm, Bayesian network algorithm, logistic regression algorithm, and Random forest classifier. All four models performed well in this setting, achieving testing accuracy between 62% and 71%.

**Table 1**

Evaluation Metrics for Selected Machine Learning Algorithms

|  | Random Forest | Bayes Network | Decision Tree | Logistic Regression |
| --- | --- | --- | --- | --- |
| Correctly Classified | 70.73% | 66.49% | 61.65% | 67.43% |
| Incorrectly Classified | 29.27% | 33.51% | 38.35% | 32.57% |
| Kappa Statistics | 0.4137 | 0.3293 | 0.2308 | 0.3484 |
| Mean absolute Error | 0.3844 | 0.3398 | 0.388 | 0.405 |
| Root mean squared error | 0.4317 | 0.5321 | 0.5959 | 0.4529 |
| Relative absolute error | 77.11% | 68.18% | 77.84% | 81.25% |
| Root relative squared error | 86.47% | 106.60% | 119.36% | 90.71% |

Random forest algorithm with featured ranking procedure was consistently the best-performing model, differentiating between trustworthy and untrustworthy users at a higher level than any of the alternatives. In addition to the overall recognition rate, we also tracked several other metrics that could describe the performance of the classifier in a more nuanced way. Some of those metrics include true and false positives rates, as well as fallout rates, ROC area measurements, recall, precision, and F1 score. All of those parameters were tracked for both samples, and were studied to understand how a classifier behaves under specific conditions.

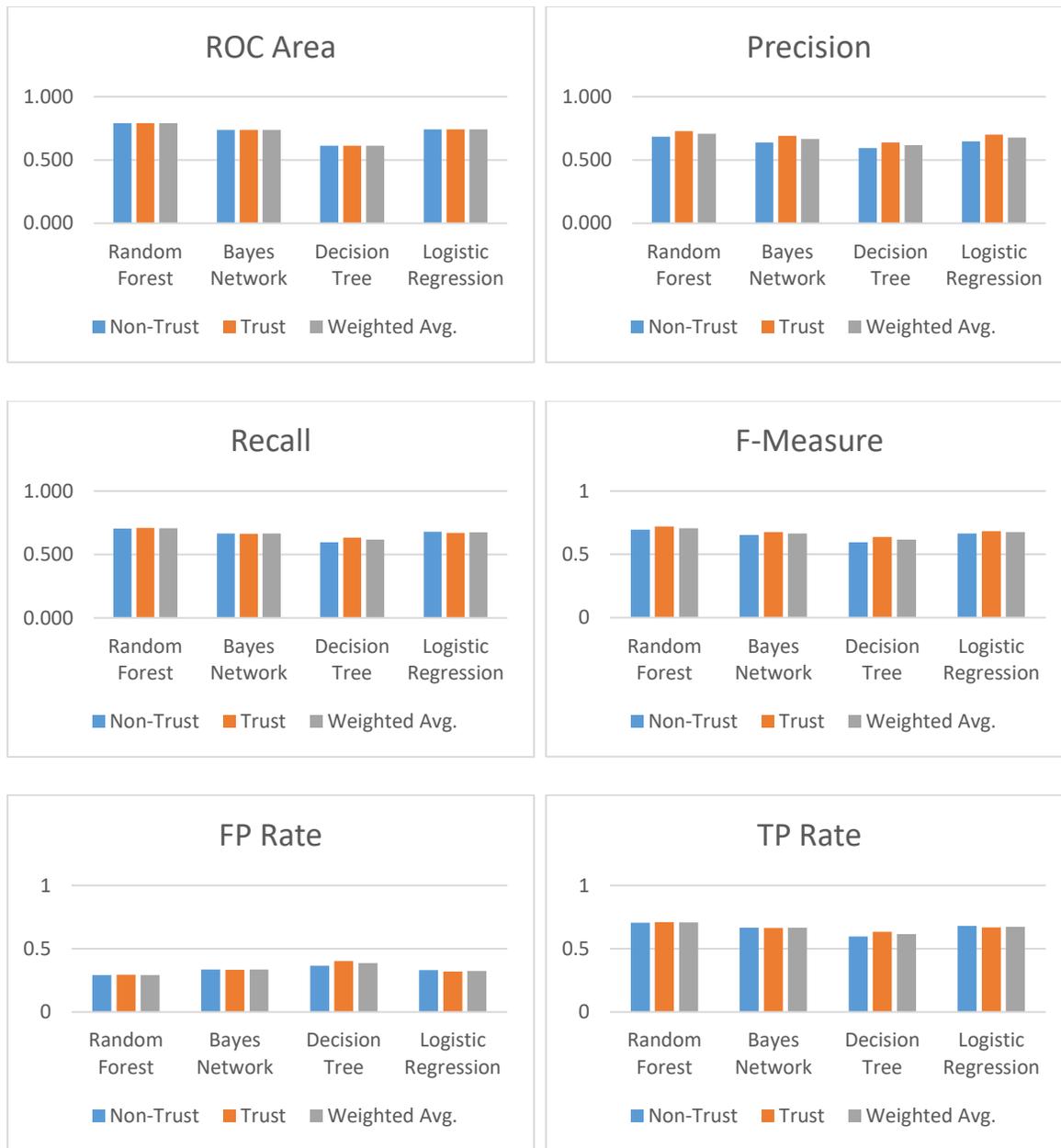

**Fig. 4.** Evaluation Metrics Charts

Most importantly, the rate of correctly recognized samples can be regarded as the model's 'responsiveness', illustrating its readiness to recognize positive examples. Conversely, false positives are basically errors, so this metric aims to determine how prone to misclassification a model may be. Those two ratios need to be well-balanced, since the objective is to create a universally valuable tool that would be able to consistently identify untrusted users without incorrectly labeling too many legitimate users. This can be seen as a special case of a binary classification task, which can be solved by setting an optimal threshold value. A good way to understand the relationship between those metrics is to use visual tools such as the ROC curve, which clearly shows how each parameter affects the ability of the solution to stay responsive without making too many errors. Another metric capable of expressing this balance between different objectives is the F1 score, which was used as a cumulative measure of performance. Using this parameter, the effectiveness of all four neural classifiers could be quantified, and they ranged from 0.617 for the decision tree method to 0.707 for the random forest model. Those results were checked for statistical significance, and it was proven they are much higher than could be explained by mere chance.

*5.7  DeepTrust Training and Evaluation*

The designed deep learning model for classification of users according to their credibility requires a training stage before it can be effectively deployed. For best results, the model should be trained on a dataset that contains numerous examples of both types of users. The collection of profiles with low credibility was compiled by establishing external references for truthfulness and identifying the authors/messages who don't abide by the truth. The possibility that otherwise completely trustworthy authors with certified profile status can post incorrect messages had to be accounted for, particularly due to the fact that good reputation of the author increases the chances that the message will be trusted and widely re-published. Because of this, all users regardless of their reputation must be examined with DeepTrust classifier. The profiles with no social connections were excluded from the sample, while a special role was reserved for individuals with relevant credentials for certain topics, as their expertise was used as a source of information for data labeling. The people tasked with labeling had previous experience with Twitter, and were instructed to search all publically available sources before making the final decision regarding truthfulness of a user profile. By gaining exposure to examples of both user types and to the constellations of various features, DeepTrust classifier gradually becomes more capable of predicting the truthfulness level of unseen samples. The training procedure was improved further by identifying the most commonly used words that describe emotional reactions and social attitudes, hence revealing additional information about the user's past activity. During the training we used a cross entropy to calculate the loss as shown in equation 5.

$$l(x,y) = \log\left(\frac{e^{x_y}}{\sum_i e^{x_i}}\right) = -x_y + \log\left(\sum_i e^{x_i}\right) \qquad (5)$$

Several hyperparameters have been utilized to minimize the loss. Learning rate α = 0.001 was tuned every run along batch normalization of 64 example for each minibatch. We also use L2 norm to minimize the overfitting and prevent the variance or biasing. The overall results of the trained/validated classifier shown in figure 3. The trained model achieves almost 97% whereas the testing accuracy is 91% and the loss is less than 10% as depicted in in figure 5.

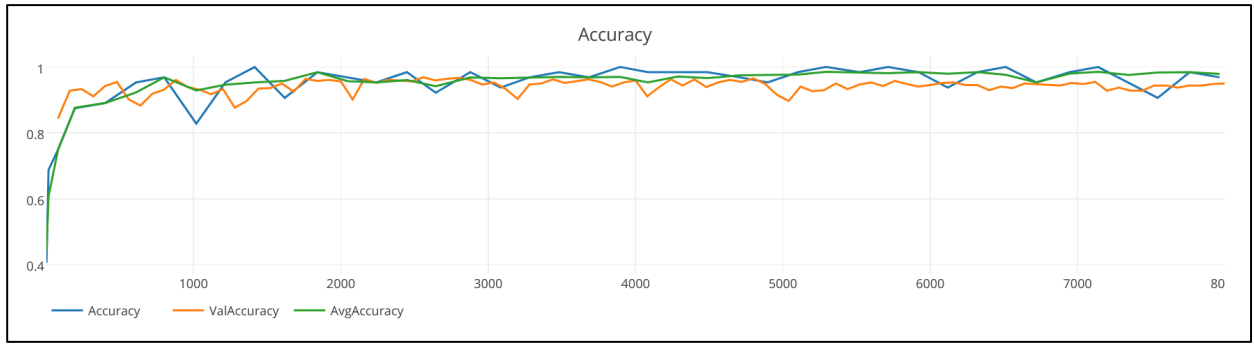

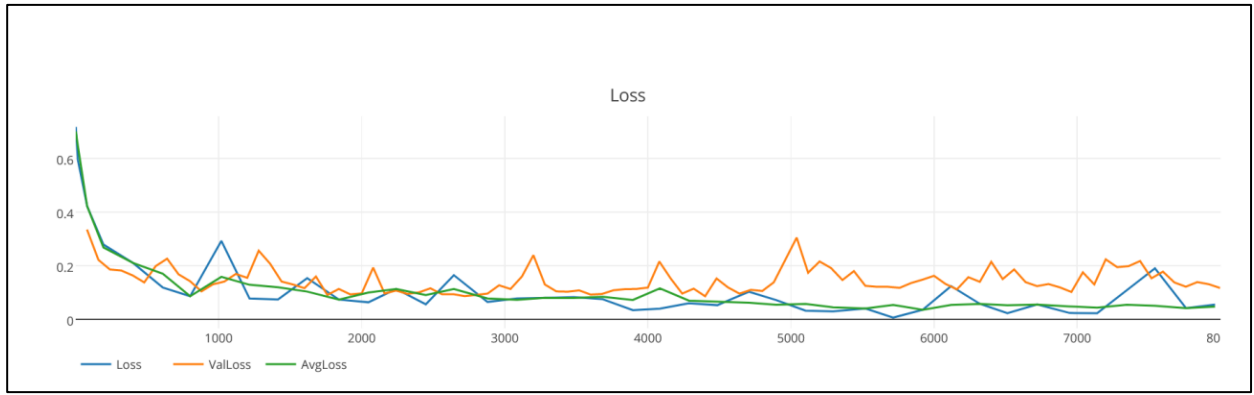

**Fig. 5.** Accuracy and Loss Metrics for DeepTrust Model

## 6  Conclusion

In this work, a new system for determining credibility of Twitter users is presented with all procedural details. The importance of this issue is compounded by the fact that a large percentage of social media users don't cross-check the information they see with reports from traditional channels. In cases when catastrophic events occur and people depend on social media reports for survival and escape, the effects of intentional or unintentional spread of information can be truly devastating.

The proposed model takes both the message and its author into account, combining the credibility assessments for each in order to obtain a more reliable measure of trustworthiness. It was tested on a dataset collected from real social networks that included more than 6.3 thousand different Twitter users. The data processing stage resulted in identification and calculation of a number of features, which were in turn utilized to train the classifier for the task of credibility evaluation. The proposed framework consists of four different modules, which are the trustworthiness classification component, user history assessment, the subjective component driven by user experience, and the ranking module. Each of those modules brings a different advantage, so user's history allows for better focusing of the model while user experience introduces human expertise into an automated system. The ranking module serves to identify the most effective features, which is essential for model optimization and further improvement of its accuracy. This concept was experimentally evaluated, and its effectiveness was confirmed on a real-world samples, albeit with a different rate of success. Those results can certainly be improved upon by refining the selection of features, and in the next stage it would be interesting to devote more attention to the factors related to temporal and spatial distribution on Twitter.

**Acknowledgment**

The authors acknowledge funding from the Research and Development (R&D) Program (Research Pooling Initiative), Ministry of Education, Riyadh, Saudi Arabia, (RPI-KSU).